\documentclass[11pt]{article}

\usepackage{amstext,fleqn,cospar}
\usepackage{url}
\usepackage{graphicx}
\usepackage[figuresright]{rotating}

\newcommand{\bez}{\begin{eqnarray*}}
\newcommand{\eez}{\end{eqnarray*}}
\newcommand{\be}{\begin{equation}}
\newcommand{\ee}{\end{equation}}
\newcommand{\beq}{\begin{eqnarray}}
\newcommand{\eeq}{\end{eqnarray}}
\newcommand{\bc}{\begin{center}}
\newcommand{\ec}{\end{center}}

\def\taut{\tau_{\rm T}}

\def\deg{\rm o}

\def\tT{\tilde{T}}
\def\tS{\tilde{S}}
\def\tH{\tilde{H}}

\newcommand{\asap}{{\it A\&A}}

\newcommand{\mnras}{{\it MNRAS}}
\newcommand{\apj}{{\it ApJ}}
\newcommand{\apjl}{{\it ApJ}}
\newcommand{\apjs}{{\it ApJS}}
\newcommand{\pasj}{{\it PASJ}}
  
\newcommand{\spscrev}{{\it Space Sci. Rev.}}  

\def\RXTE{{\it RXTE}}

\def\EXOSAT{{\it EXOSAT}}

\newbox\grsign \setbox\grsign=\hbox{$>$} \newdimen\grdimen \grdimen=\ht\grsign
\newbox\simlessbox \newbox\simgreatbox \newbox\simpropbox
\setbox\simgreatbox=\hbox{\raise.5ex\hbox{$>$}\llap
     {\lower.5ex\hbox{$\sim$}}}\ht1=\grdimen\dp1=0pt
\setbox\simlessbox=\hbox{\raise.5ex\hbox{$<$}\llap
     {\lower.5ex\hbox{$\sim$}}}\ht2=\grdimen\dp2=0pt
\setbox\simpropbox=\hbox{\raise.5ex\hbox{$\propto$}\llap
     {\lower.5ex\hbox{$\sim$}}}\ht2=\grdimen\dp2=0pt
\def\ga{\mathrel{\copy\simgreatbox}}
\def\la{\mathrel{\copy\simlessbox}}

\title{UNDERSTANDING
SPECTRAL VARIABILITY AND TIME LAGS IN ACCRETING BLACK HOLES}

\author{Juri Poutanen \address{Stockholm Observatory,
    SE-133 36 Saltsj\"obaden, Sweden}}

\begin{document}

\maketitle

\begin{abstract}
I review the temporal/spectral  data of accreting black hole sources paying
most attention to the properties of the temporal variability such as photon
energy  dependent  auto-  and  cross-correlation  functions,  average  shot
profiles  and  hardness  ratios,  and  the  Fourier   frequency   dependent
time/phase lags.  These statistics  characterize  spectral changes at short
time  scales  that are  otherwise  impossible  to study by direct  spectral
analysis.  The data provide strong  constraints on the  theoretical  models
for X-ray  production  in accreting  black  holes.  Models for the spectral
variability of the Comptonized  component are briefly reviewed.  It is also
shown that Compton  reflection can have significant  impact on the observed
temporal characteristics.
\end{abstract}

\section*{INTRODUCTION}

Accreting  Galactic black holes (GBHs) in X-ray  binaries and  supermassive
black  holes in active  galactic  nuclei  (AGNs) in the  centers of Seyfert
galaxies  show  similar  X-ray/$\gamma$-ray  spectra.  In the  X-ray  band,
spectra can be described as a cutoff  power-law  with a Compton  reflection
bump atop of that (see  Fig.~1a).  All GBHs in their hard states  that were
observed  with  sufficient   sensitivity  above  100  keV  (e.g.  by  Sigma
instrument onboard the {\it Granat Observatory} and OSSE instrument onboard
the {\it Compton Gamma-Ray  Observatory})  show a sharp cutoff at about 100
keV (e.g.  Gilfanov et al.  1995; Grebenev et al.  1993, 1997; Grove et al.
1998a;  Gierli\'nski et al.  1997; Zdziarski et al.  1997; Zdziarski 1999).
Similar cutoff is observed in the spectra of bright  Seyfert  galaxies such
as NGC 4151 and NGC 5548 (Zdziarski,  Johnson, \& Magdziarz 1996; Magdziarz
et al.  1998), and in the composite spectrum of a number of weaker Seyferts
(Zdziarski, Poutanen, \& Johnson 2000).

It is broadly  accepted  that the only  mechanism  capable of  producing  a
power-law  spectrum  with the cutoff at a similar  energy in  objects  that
differ by orders of magnitude in luminosity, size, and a BH mass is thermal
Comptonization (e.g.  Sunyaev \& Titarchuk 1980).  The observed spectra are
produced in a cloud of hot electrons with the temperature $kT_e\sim 50-100$
keV and Thomson optical depth  $\taut\sim 1$ (for a review see Zdziarski et
al.  1997).  Only in the brightest  objects such as Cyg X-1 deviations from
thermal Comptonization model at high energies are statistically significant
(McConnell  et al.  2000b).  An excess of the emission  above $>1$ MeV (see
Fig.~1b) is a possible  signature of  non-thermal  electrons  in the source
(Li, Kusunose, \& Liang 1996; Poutanen 1998).  The  X/$\gamma$-ray  spectra
of GBHs in the soft state are  dominated  by a black body type  emission at
$\sim 1$ keV with a power-law  tail  extending  at least up to 400-800  keV
(Grove et al.  1998a; Gierli\'nski et al.  1999) or maybe even up to 10 MeV
(McConnell  et  al.  2000a).  The  tail  is  most   probably   produced  by
Comptonization in non-thermal plasmas (Poutanen \& Coppi 1998; Gierli\'nski
et al.  1999; for reviews see Poutanen 1998; Coppi 1999).

\begin{figure}[t]
\begin{minipage}{9.cm}
\includegraphics[width=85mm,height=65mm]{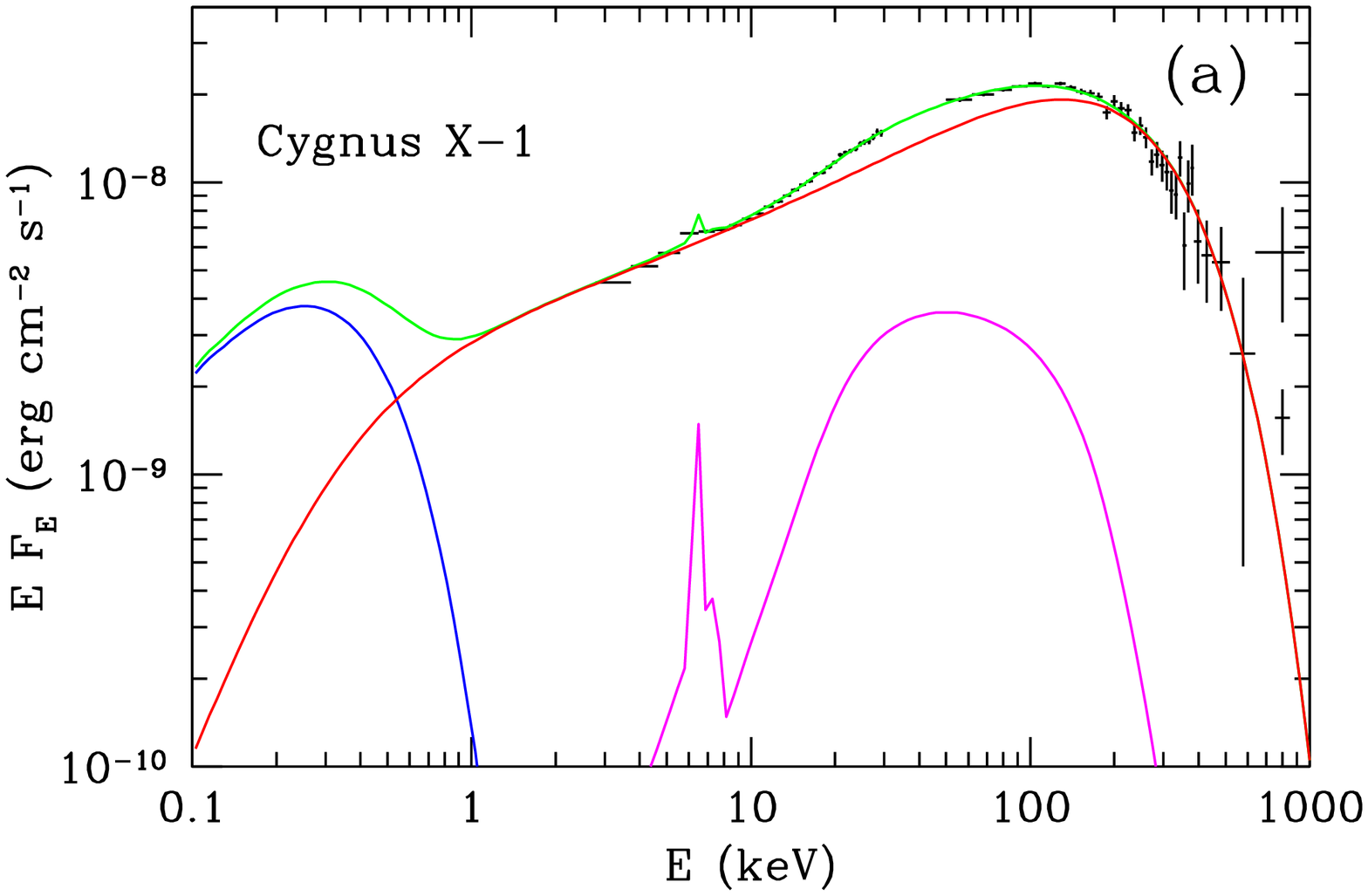}
\end{minipage}
\begin{minipage}{9.cm}
\includegraphics[width=85mm,height=65mm]{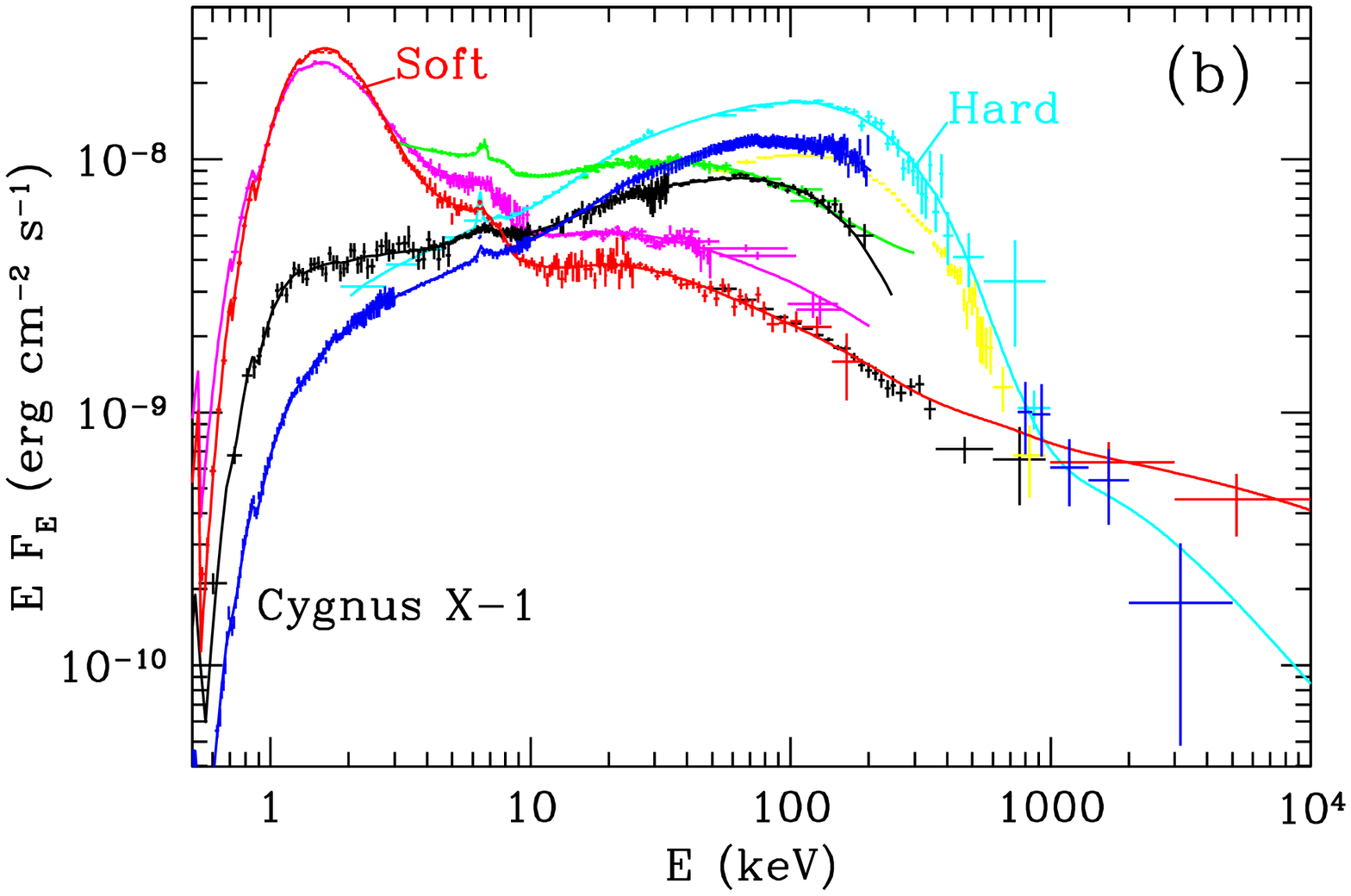}
\end{minipage}


{\sf Fig. 1. (a) Cygnus X-1 in the hard state (data from 
Gierli\'nski et al. 1997).
Spectrum can be decomposed into several components: thermal Comptonization,
Compton reflection, and a soft excess (modeled here as a $\sim 0.1$ keV 
black body).
(b) Spectral states of Cygnus X-1. Courtesy of A.~A. Zdziarski.}
\end{figure}

All spectral models  (particularly those incorporated into {\sc xspec}) are
designed to fit the time averaged spectra.  However, it is known that black
holes show spectral variability  reflected in time lags, asymmetries of the
cross-correlations  function, and changes of the hardness ratio.  On longer
time scales  (minutes-hours  for GBHs, days for AGNs), spectral changes can
be  tracked  directly  by  fitting  the  spectrum  averaged  over that time
interval and following the changes in the best fit parameters.  However, it
is not possible to  determine  directly  how the  spectrum  changes on very
short time scales  corresponding to the characteristic time scales close to
the  black  hole.  One can do it  only  in a  statistical  sense,  i.e.  by
analyzing different statistics of the  temporal-spectral  variability, such
as auto- and  cross-correlation  functions  (ACF/CCF) in  different  energy
channels,  power-density  spectra (PDS),  cross-spectra,  phase/time  lags,
coherence  function,  etc.  that contain  important  information  about the
physical processes responsible for X-ray production.

Ideally, any model describing the spectral properties of black holes should
also describe their temporal characteristics.  In this review, I will first
discuss the data obtain by time domain analysis.  Then the results obtained
via Fourier analysis are presented.  Finally, I will review models proposed
to describe temporal and spectral properties.

\section*{VARIABILITY IN THE TIME DOMAIN}

\subsection*{Auto- and cross-correlation functions of GBHs}

Time  domain  techniques  were  used at the very  early  days of the  X-ray
astronomy.  In the 70-ies, the auto- and cross-correlation  function of Cyg
X-1  were  obtained  (Weisskopf,  Kahn,  \&  Sutherland  1975;  Sutherland,
Weisskopf,  \& Kahn 1978;  Priedhorsky  et al.  1979;  Nolan et al.  1981).
The  asymmetry  of the CCF of Cygnus X-1 was  discovered  in $\sim 150$ sec
observations  by Priedhorsky et al.  (1979) and Nolan et al.  (1981).  They
showed  that the CCF peaks at a lag $\la  10-40$  ms.  Using  the  \EXOSAT\
data, Page (1985)  confirmed  these results and claimed a $\sim 6$ ms shift
of the peak of the CCF  between  the  5-14 keV and the 2-5 keV  bands.  For
many  years, the time  domain  analysis  was not  applied  to GBHs  despite
immense advances in temporal resolution, photon statistics, and duration of
observations.  Aside from  attempts  to model  individual  shots  (Lochner,
Swank, \& Szymkowiak  1991; Negoro et al.  1994; Feng et al.  1999), recent
analyses have concentrated on Fourier domain techniques.

Recent \RXTE\  observations  (Maccarone,  Coppi, \& Poutanen  2000) clearly
show  asymmetries of the CCFs, which, however, peak within $\sim 1$~ms from
zero lag  (see  Fig.~2).  This  suggests  that  the  relation  between  the
variation in the two bands are not simply a time delay.  The rising part of
the CCF (soft lags) becomes narrower with energy substantially  faster than
the decaying part (hard lags).  This is in qualitative  agreement  with the
results of  Priedhorsky  et al.  (1979) and Nolan et al.  (1981).  The CCFs
reach  values  very  close to unity,  showing  that the  signal  at all the
energies is extremely well synchronized.  Asymmetry is also observed in the
\RXTE\ data of GX 339-4,  where the CCFs are offset by $\la 5$~ms from zero
(using the 2-5 and 10-40 keV bands, Smith \& Liang 1999).  The CCFs of AGNs
also display similar  properties  (e.g., Papadakis \& Lawrence 1995; Lee et
al.  2000).

Maccarone  et al.  (2000)  presented  also the ACFs of Cygnus X-1 which are
shown in  Figures~2  and 3a.  The width of the ACF  decreases  with  photon
energy  approximately as $\propto E^{-0.2}$ at lags smaller than $\sim 0.3$
sec.  (At larger lags the ACFs at different energies are not self-similar.)
This strongly  constrains the origin of the spectral  variability, since it
requires  that the pulses  producing  the  variability  last  longer at low
energies than at higher energies.


\begin{figure}[t]
\bc
\begin{minipage}{13.cm}
\includegraphics[width=140mm,height=100mm]{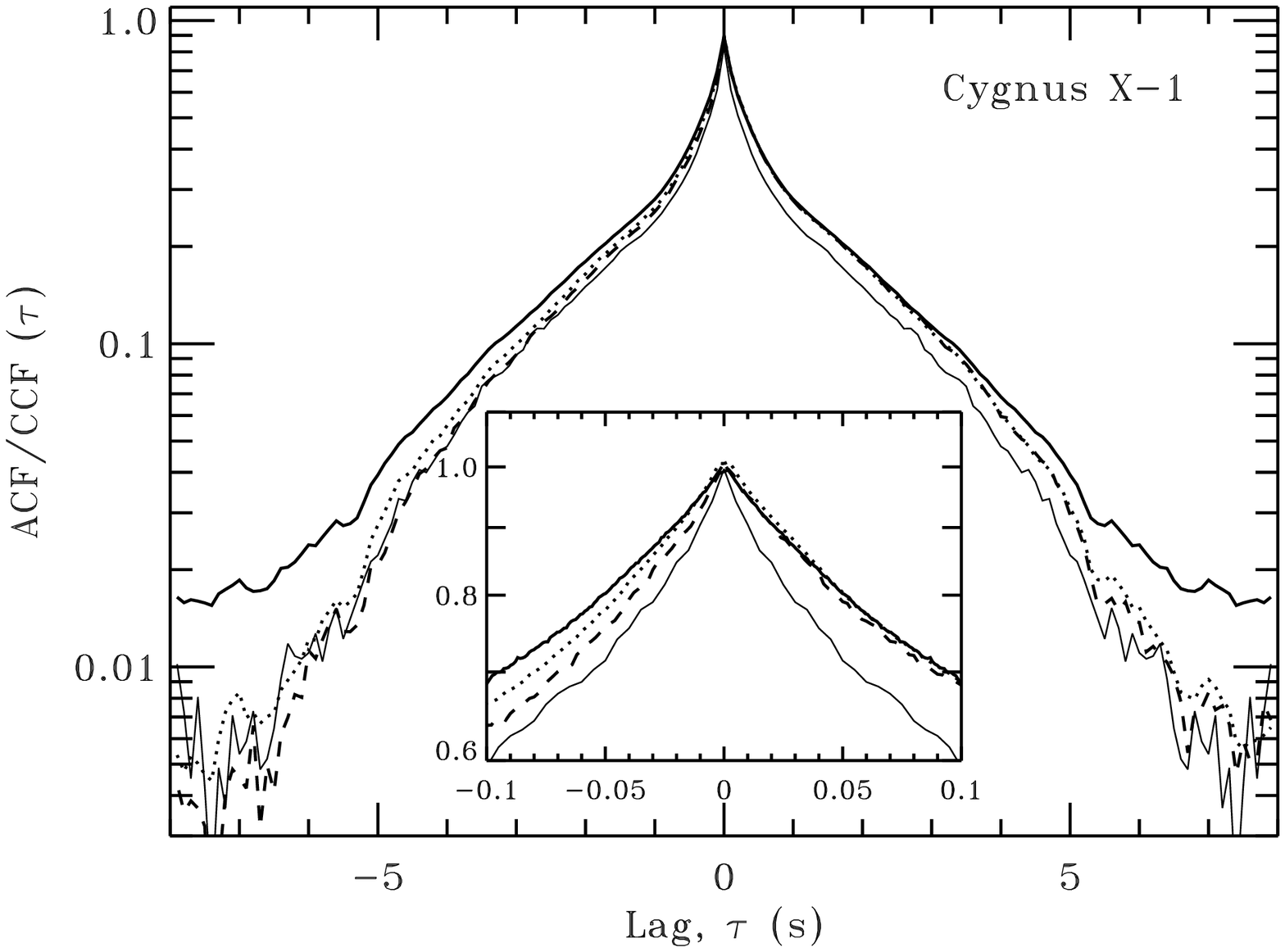}
\end{minipage}
\ec


{\sf Fig. 2. 
The auto-  and  cross-correlation  functions  of  Cygnus  X-1  observed  in
December  1997.  Solid  curves  show the ACFs for the 2-5 keV  energy  band
(thick  curves) and the 24-40 keV band (thin  curves).  Dotted  curves show
the CCF  between the 8-13 keV band and the 2-5 keV band, and dashed  curves
represent  the CCF for the 24-40 keV band vs the 2-5 keV  band.  The  peaks
all align at around zero lag.  The higher energy curves are  narrower.  The
CCFs are  defined  in such a way that the peak is  expected  to appear at a
positive lag when hard photons are lagging the soft ones.  From  Maccarone,
Coppi, \& Poutanen (2000).
}
\end{figure}

\mbox{ }

\begin{minipage}{9.cm}
\includegraphics[width=87mm,height=70mm]{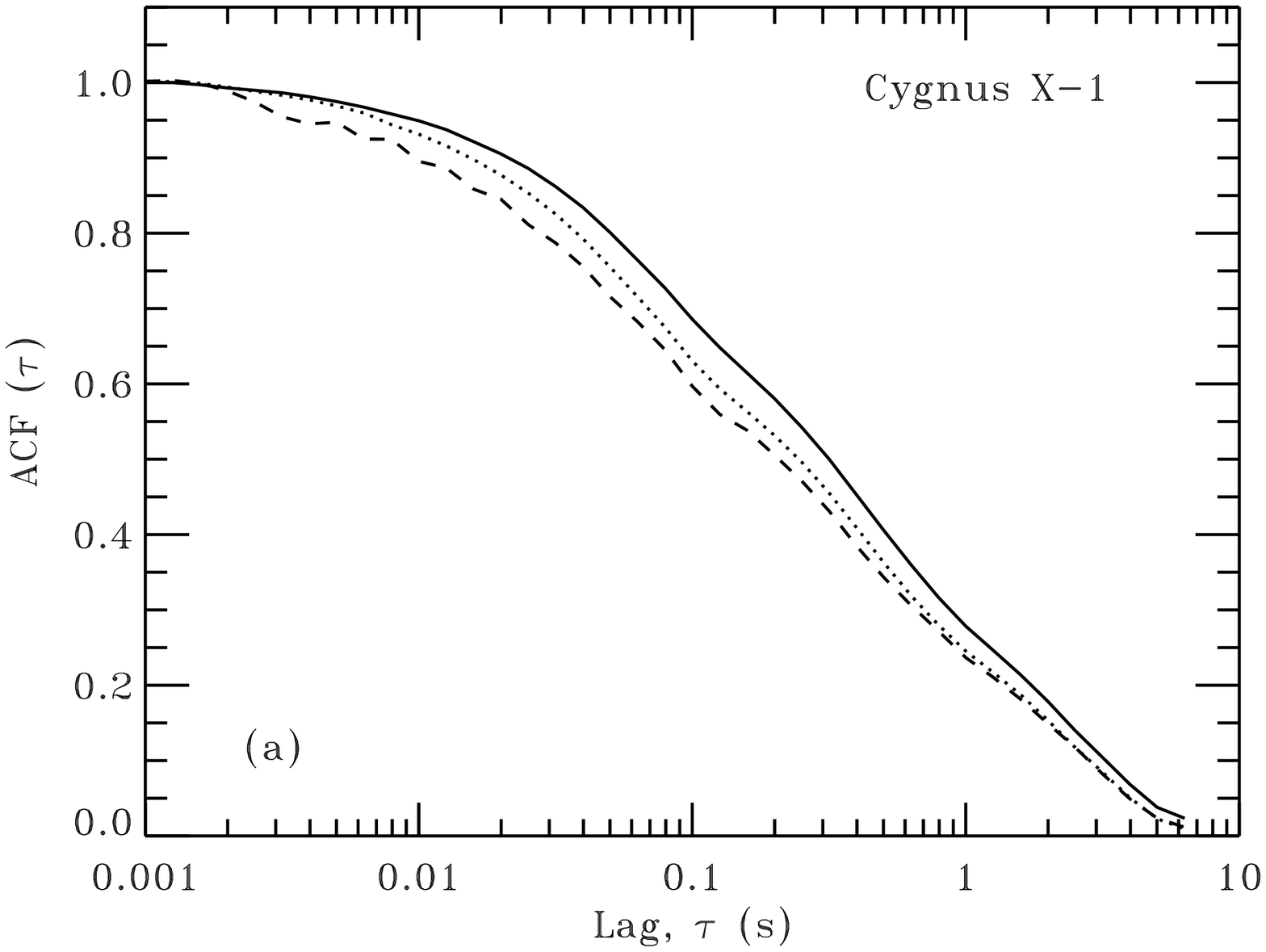}
\end{minipage}
\begin{minipage}{9.cm}
\includegraphics[width=87mm,height=70mm]{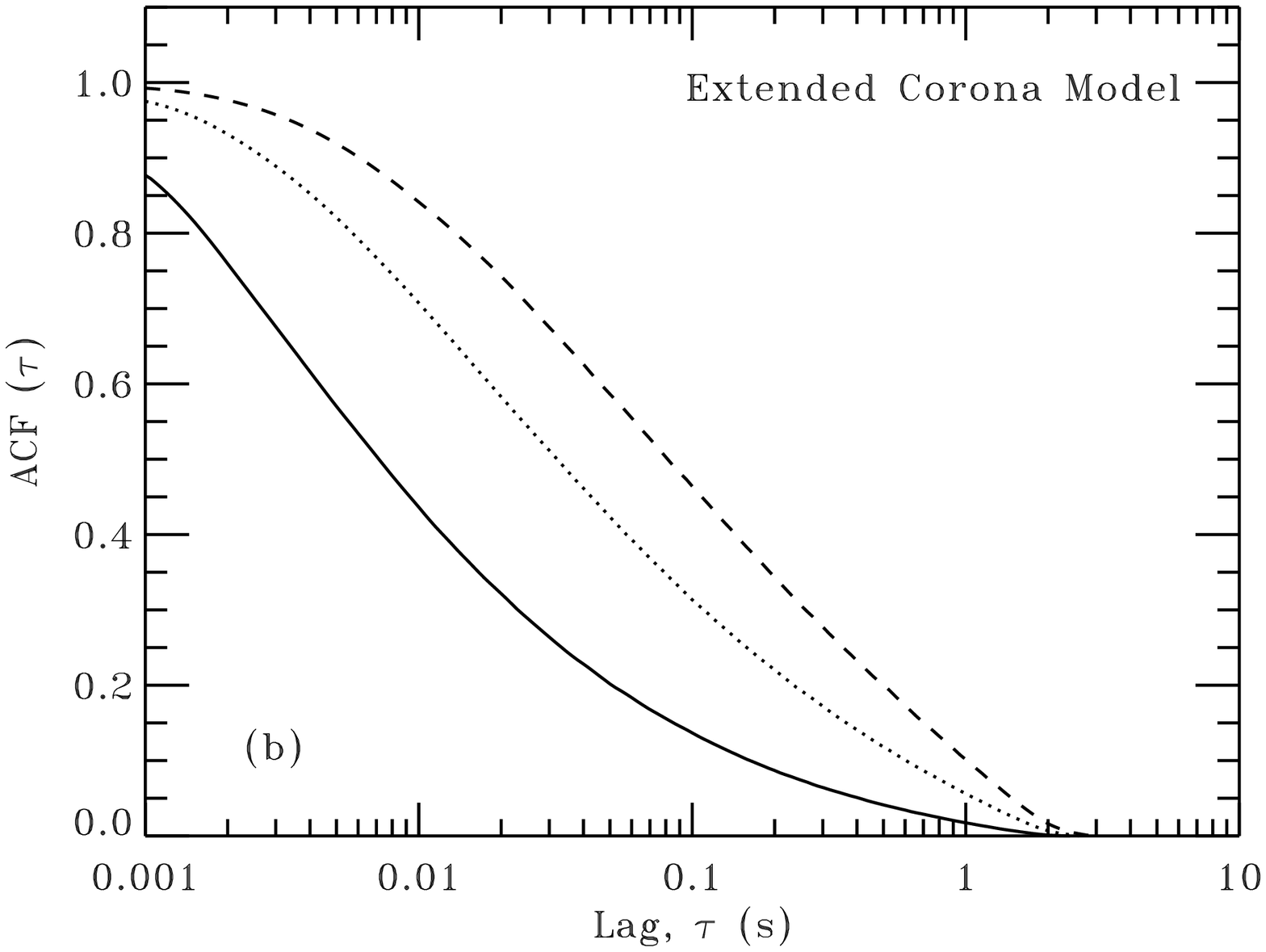}
\end{minipage}


{\sf Fig. 3. The ACFs in the  different  energy  bands.  
(a) The data  for Cygnus  X-1.
Solid curves shows the ACF for the 2-5 keV energy band, dotted curve is for
8-13 keV band, and dashed  for 24-40 keV band.  
(b) The  prediction  of the extended corona model (Kazanas et al. 1997;
Hua et al. 1999). 
From Maccarone et al. (2000). }


\begin{figure}[hpbt]
\begin{minipage}{9.cm}
\includegraphics[width=40mm,height=80mm,angle=270]{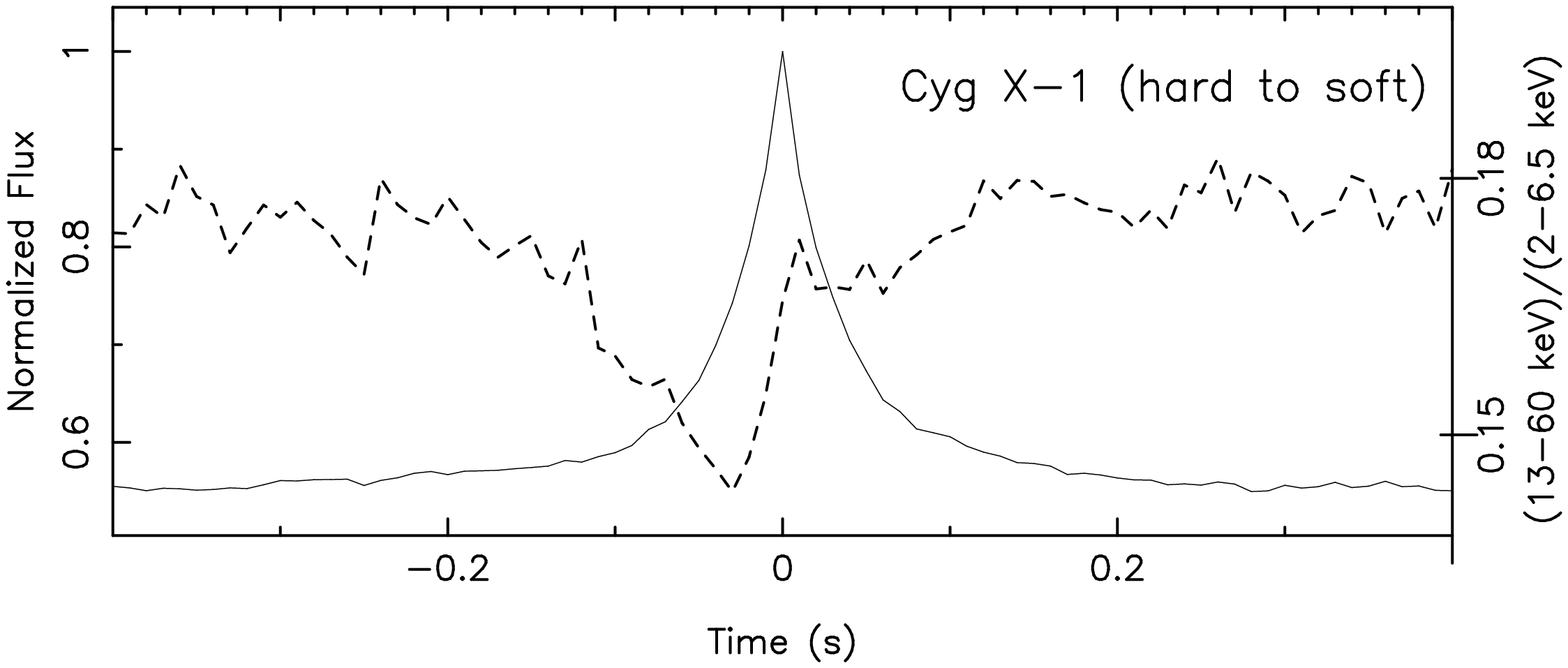}
\includegraphics[width=40mm,height=80mm,angle=270]{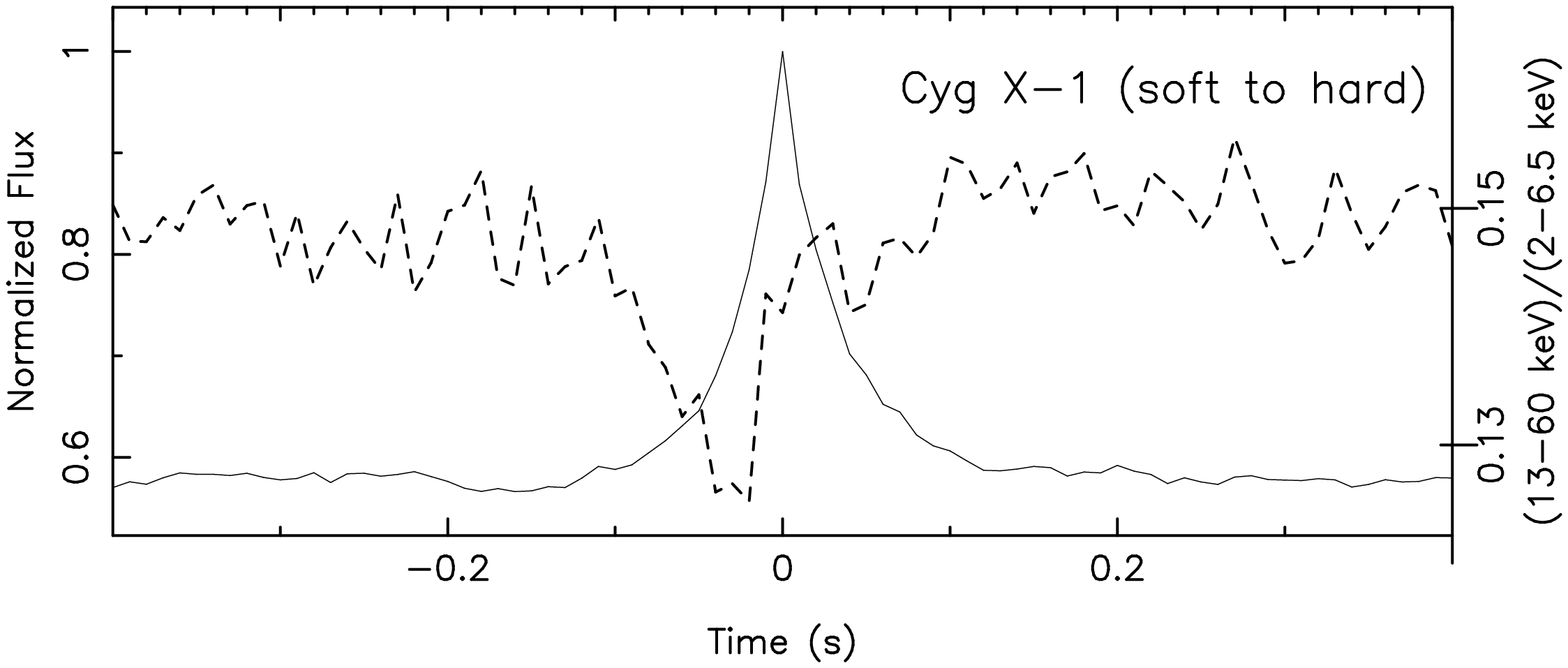}
\end{minipage}
\begin{minipage}{9.cm}
\includegraphics[width=40mm,height=80mm,angle=270]{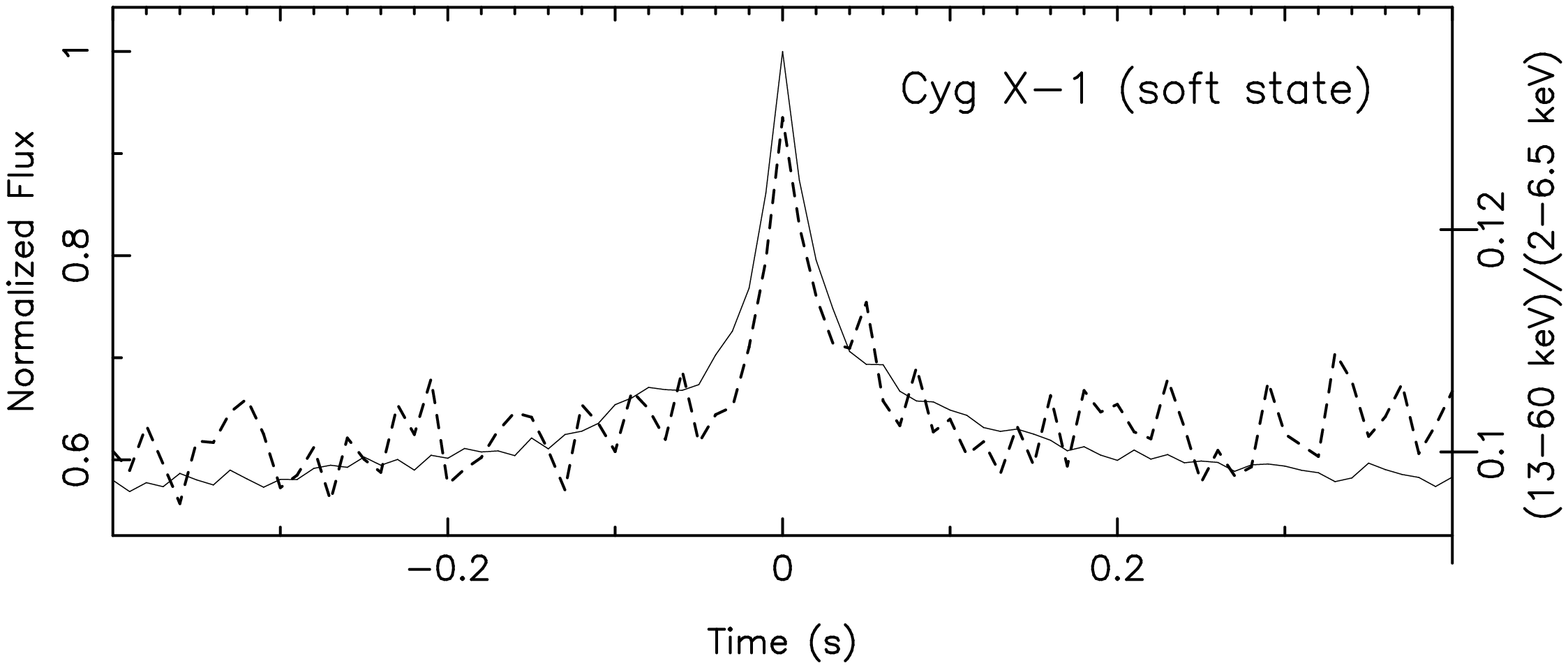}
\includegraphics[width=40mm,height=80mm,angle=270]{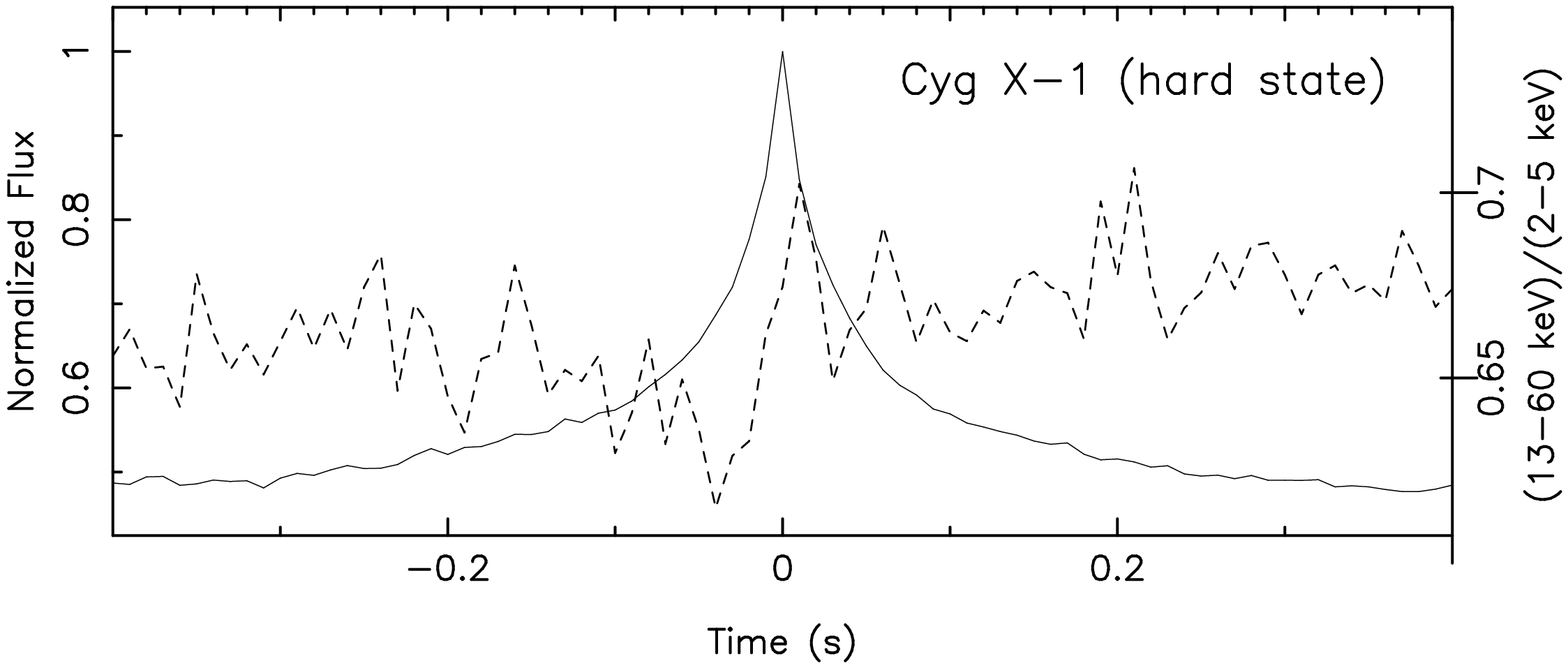}
\end{minipage}


{\sf Fig. 4. Peak aligned average shots in Cygnus X-1
and their hardness ratios.
In different spectral states the hardness (dashed curves)
varies   differently along the shot light curve (solid curves).
From Li, Feng, \& Chen (1999).}
\end{figure}

\newpage

\subsection*{Average Peak Aligned Shots}

A different approach to analyzing  variability in the time domain was taken
by Negoro,  Miyamoto, \& Kitamoto (1994), Feng, Li, \& Chen (1999), and Li,
Feng, \& Chen (1999).  They  constructed the average peak aligned  profiles
of the shots in Cyg X-1 at different energies (see Fig.~4).  This procedure
is rather  similar  to that used in the  analysis  of the  light  curves of
gamma-ray  bursts (see e.g.  Stern,  Poutanen, \& Svensson 1997, 1999).  In
spite of the fact that  definition  of a shot itself is rather ad hoc, this
procedure  provides some insight into spectral  variability  during  bright
emission  episods.  In the hard state, the shot becomes  first  softer, and
then, just after the peak, harder than the average emission  (Negoro et al.
1994).  In the soft state, shots are harder, and in the  transition  states
softer than the time average emission.

The behavior in the soft state is easy to understand.  It is known that the
disk (black body like) emission dominating the 2-6 keV energy band does not
vary much, while the  Comptonized  hard  (power-law  like) tail  dominating
energy  band above 13 keV is  strongly  variable  (Churazov,  Gilfanov,  \&
Revnivtsev  2001).  An increase in flux  (``shot'')  implies an increase in
the  normalization of the hard tail and larger hardness ratio.  It would be
of interest to see,  however,  how the  hardness in the energy  bands above
$\sim 10$ keV  varies.  This  would  provide  important  information  about
spectral variability of the hard tail.

The  transition  state  of Cyg X-1 is  characterized  by  strong  QPO  type
oscillations  at $\sim  1-10$ Hz.  The  softening  of the  spectrum  during
increase of the count rate can be a result of those oscillations that cause
changes in the  geometry of the  emission  region,  e.g.  by  changing  the
overlap  between  hot inner disk and cold outer  disk (see  Fig.~5a).  Such
changes  can be similar to the long term  changes  characterizing  spectral
transitions  (Poutanen,  Krolik,  \&  Ryde  1997;  Esin  et al.  1998;  see
Fig.~1b).  More overlap  increase  Compton cooling by soft photons from the
cold disk, causing  temperature of the Comptonizing  region to decrease and
softening the spectrum.  In the magnetic  flare model with plasma  ejection
(Beloborodov  1999a,b),  similar  spectral  evolution  occurs if the plasma
ejection  velocity is variable.  Smaller  velocity means more feedback from
the cold disk, more soft seed photons  coming to the  emission  region, and
softer  spectrum.  One should caution the reader that at energies  above 13
keV most of the  variability  power  emerges  at  $\sim  10$ Hz QPO and the
average  shot  techniques  does  not say  almost  anything  about  spectral
evolution at these time scales.

The  hard-state  spectral  evolution  is the  most  complicated  one.  The
observed  behavior could be reproduced by a magnetic flare model (Poutanen
\&  Fabian  1999)  where  increasing  separation  of  the  flare  from  the
underlying  cold disk  produces  changes in the feedback and  corresponding
spectral evolution~(Fig.~5b).

\begin{figure}[t]
\begin{minipage}{9.2cm}
\includegraphics[width=90mm,height=75mm]{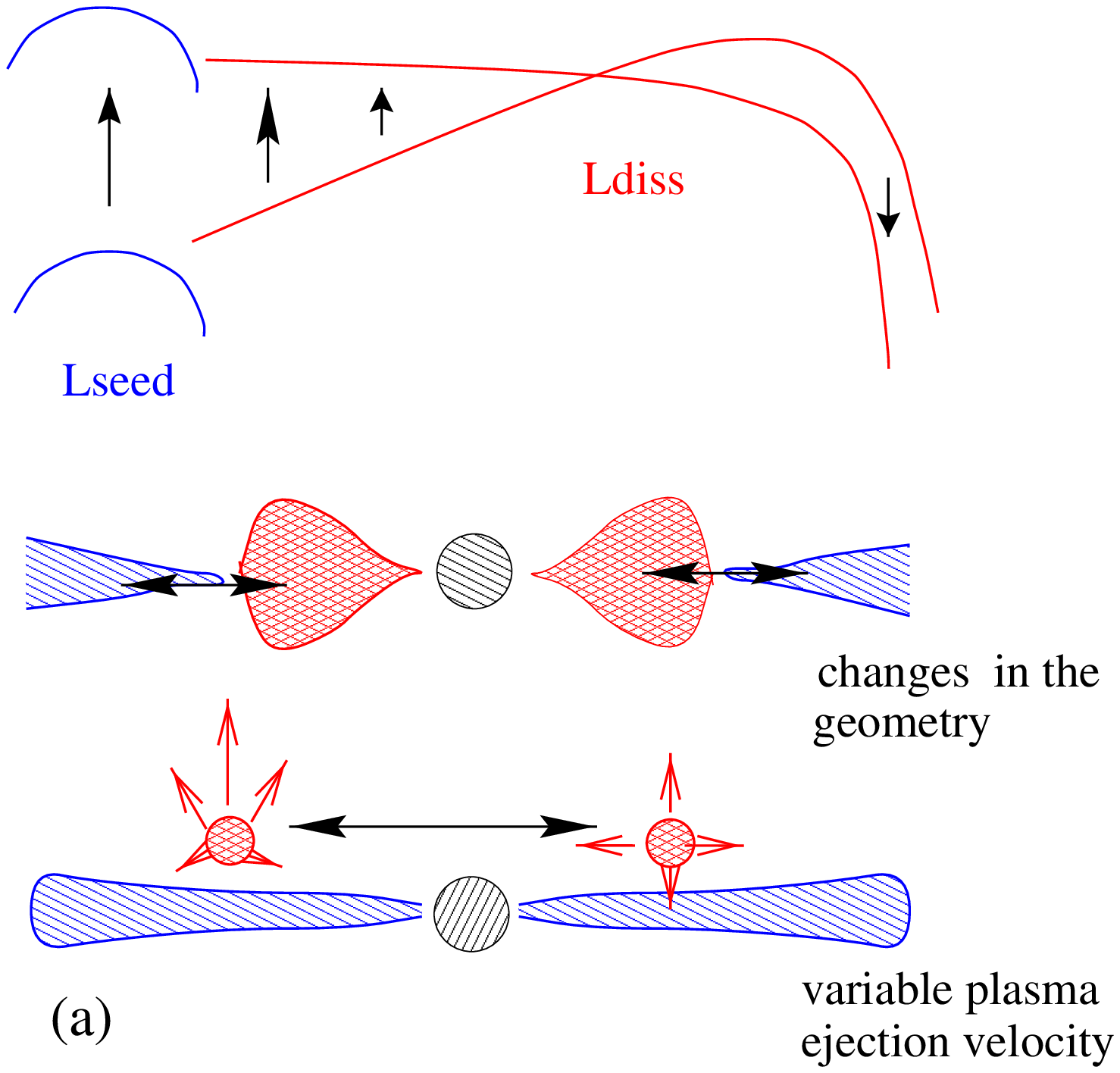}
\end{minipage}
\begin{minipage}{8.8cm}
\includegraphics[width=77mm,height=75mm]{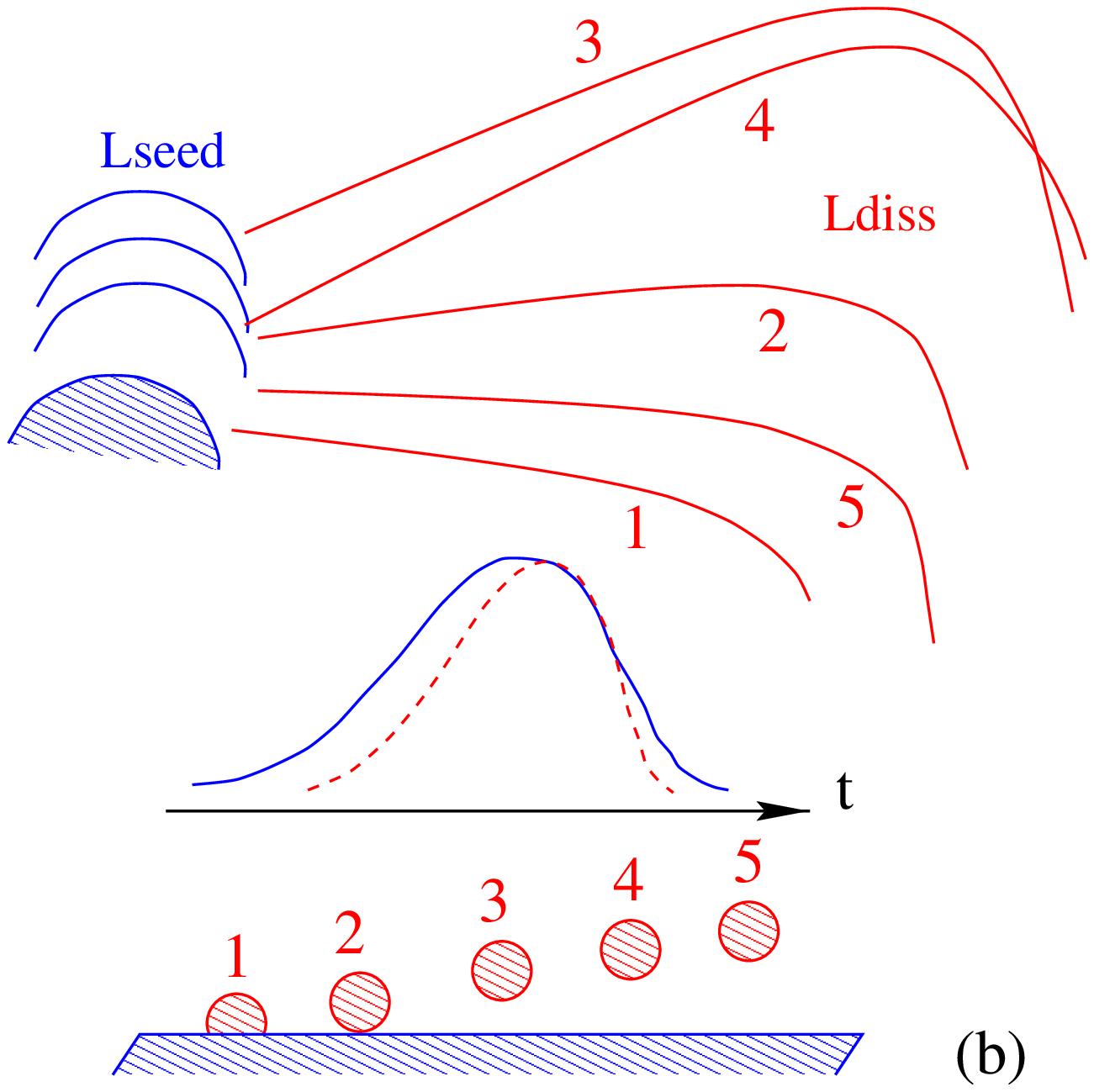}
\end{minipage}


{\sf Fig. 5. 
Cartoons of the variability  mechanisms.  (a)  Changes in the geometry or
ejection  velocity  cause  corresponding  changes  in the seed soft  photon
luminosity  entering  the hot  Comptonizing  plasma  cloud.  
The spectrum  pivots if the heating  rate in the hot cloud stays constant. 
Such variability  could be responsible for the observed  spectral changes 
in the transition state.
(b)  Spectral evolution of the magnetic flare (Poutanen \& Fabian 1999).
Movement of the X-ray emitting cloud from the  underlying  disk changes the
feedback.  This  in  turn  causes  soft-hard-soft  evolution  of the  flare
spectrum that satisfies observational  constraints in the hard state of Cyg
X-1. If the energy dissipation rises slower than it decays,   
shots at higher energy (dashed curve) are narrower than low energy 
shots (solid curve) satisfying constraints from the ACF (see Fig. 3).
}
\end{figure}

\section*{FOURIER ANALYSIS AND TIME LAGS}

In spite of the fact that Fourier domain  functions  contain, in principle,
the same  information  as their  time  domain  companions,  they  highlight
different  information  which can be used to constrain  the models.  During
last decade a number of authors studied  Fourier-frequency--dependent  time
lags between photons in different  energy channels for many accreting black
holes and neutron  stars (see e.g.  Miyamoto et al.  1988, 1992; Cui et al.
1997; Nowak et al.  1999a,b; and  Poutanen  2001 for a recent  review).  By
computing the  cross-spectrum  $C(f)\equiv  \tS^*(f) \tH(f)$ (here $\tS(f)$
and $\tH(f)$ are the Fourier transforms of the light curves in the soft and
hard energy channels,  respectively, and $f$ is the Fourier frequency), the
phase lag, $\delta\phi(f)\equiv \arg[C(f)]$, can be obtained.  The time lag
is then, $\delta  t(f)\equiv\delta\phi(f)/(2\pi f)$.  The lags are positive
when hard photons are lagging the soft ones (hard lags).

There are a few  important  facts to notice.  First, time lags are normally
hard (except  when strong  quasi-periodic  oscillations  are present in the
data, see e.g.  Cui 1999;  Reig et al.  2000 for the case of GRS  1915+105,
and Cui, Zhang, \& Chen 2000 for the case of XTE  1550-564).  Second,  they
depend  approximately   logarithmically  on  photon  energy,  i.e.  $\delta
t(f)\propto \ln  (E_h/E_s)$,  where $E_h$ and $E_s$ and the energies of the
``hard'' and ``soft'' channel (Miyamoto et al.  1988; Nowak et al.  1999a).
Third, time lags depend on the  Fourier  frequency  in a quite  complicated
fashion (see Fig.  6a).  At high frequencies, $\delta t(f)\propto 1/f$.  It
is also  interesting  to note that in Cyg X-1 as well as in GX 339-4 (Nowak
et al.  1999b) and GRO  J0422+32  (Grove et al.  1998b) there is a break in
the time-lag spectrum at $f\sim 0.1-0.5$ Hz.

\begin{figure}[t]
\begin{minipage}{10cm}
\includegraphics[width=100mm,height=80mm]{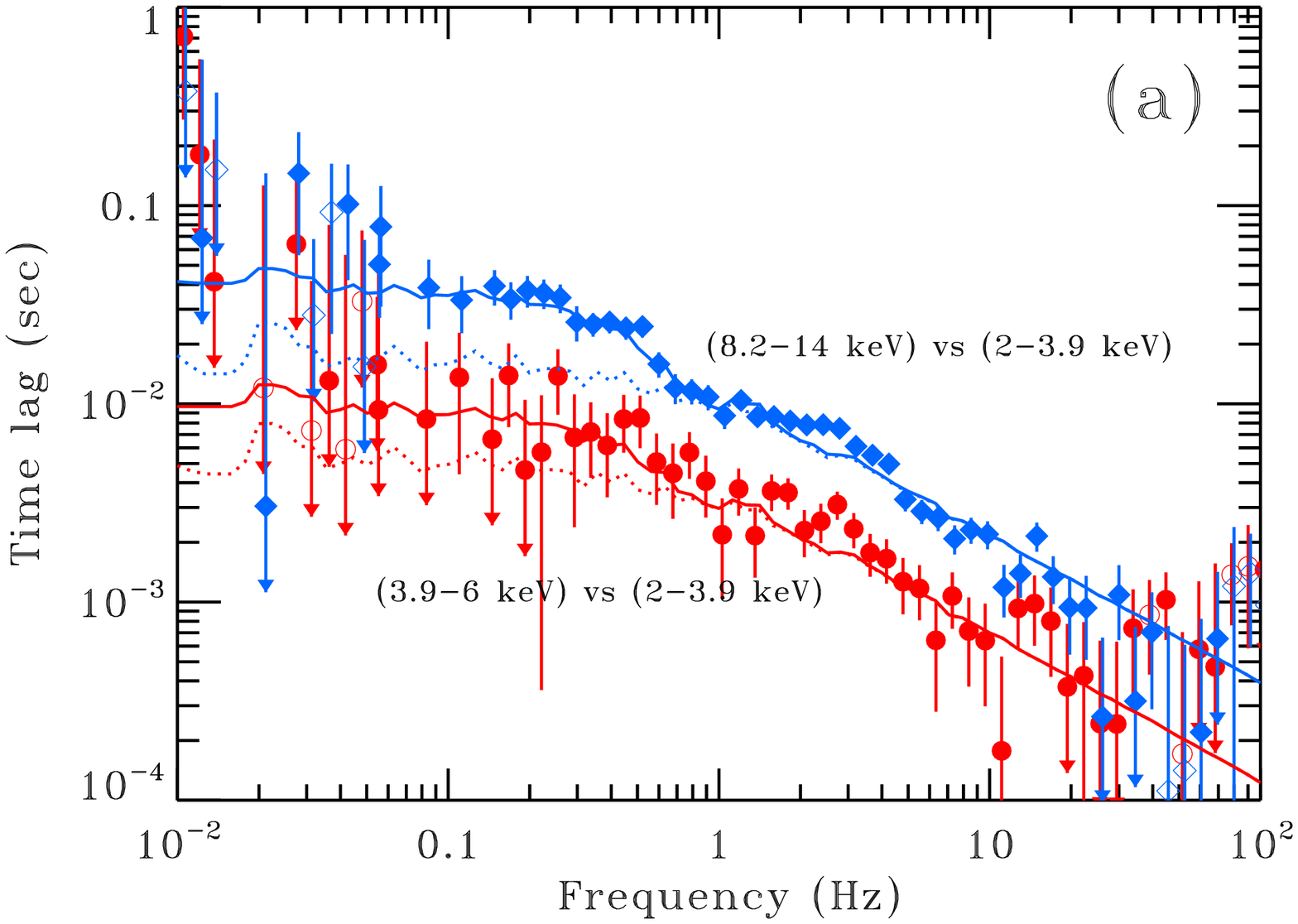}
\end{minipage}
\begin{minipage}{8cm}
\includegraphics[width=80mm,height=110mm]{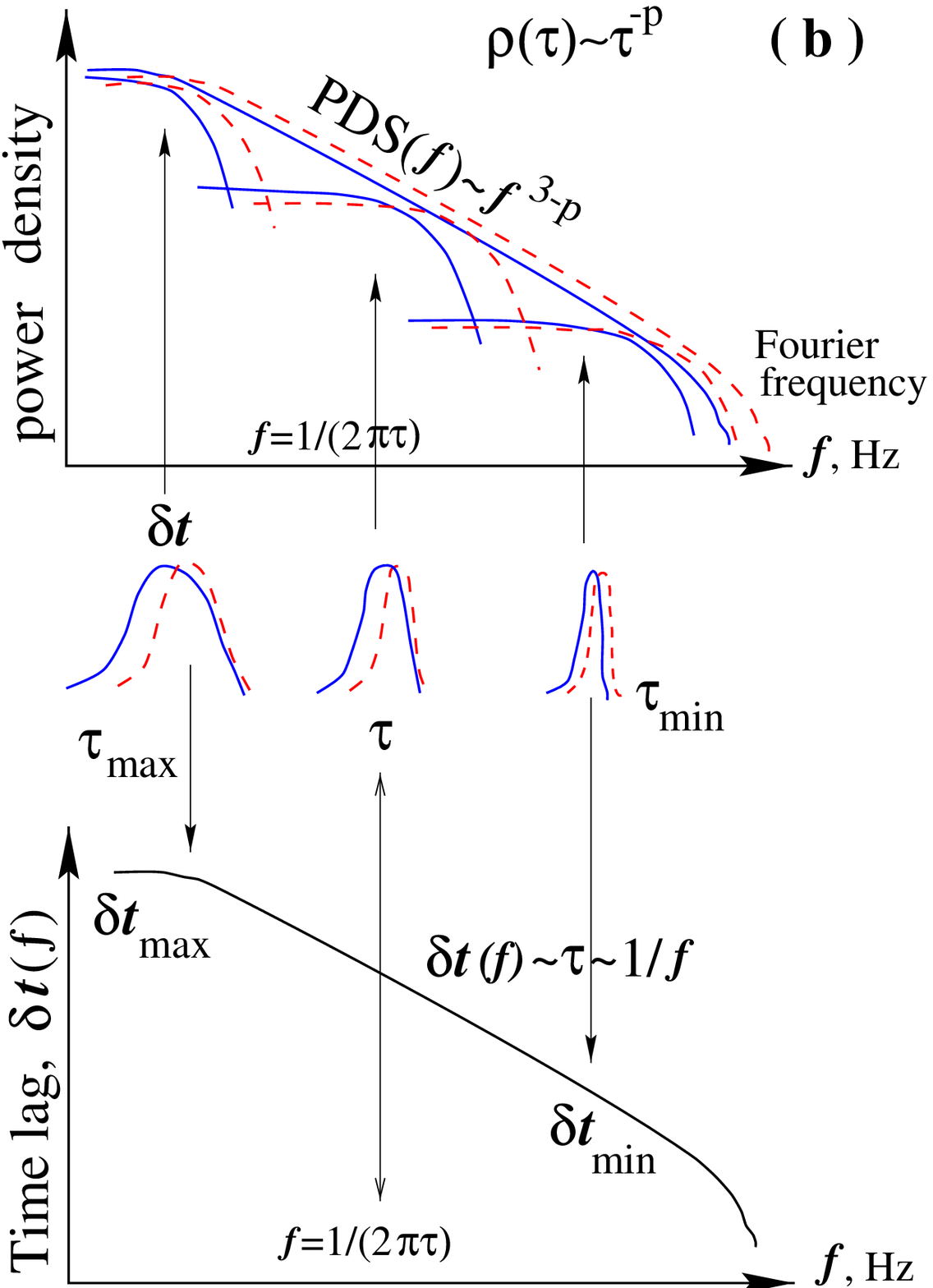}
\end{minipage}

{\sf Fig. 6.   
(a) Time lags  in the hard  state of Cyg X-1 (Nowak et al. 1999a;
Pottschmidt  et al.  2000).  Dotted curves show the modified  shot
noise model with time scales  between  $\tau_{\min}=1$~ms  and
$\tau_{\max}=0.2$~s   (Poutanen  \&  Fabian  1999).  The  shots  at  higher
energies  are delayed as shown on the right  panel (see also  Fig.~5b).   
A fraction of hard photons can be delayed  due to Compton reflection
in the  outer  part  of  the  disk,  time  lags  increase then at  low  
Fourier frequencies. The total lags are shown by solid curves.  \\
(b) Time lags and PDS for a  modified  shot  noise  model.  Shots at softer
energies  are shown by solid  curves,  and at  higher  energies  by  dashed
curves.  For a power-law  distribution  of time  scales  $\rho(\tau)\propto
\tau^{-p}$,  the  resulting  PDS is a power-law.  If the delay  between the
shots at  different  energies  is a  constant  fraction  of  $\tau$,  the
resulting time lags are $\delta t(f)\propto 1/f$.
}
\end{figure}

\section*{MODELS FOR SPECTRAL VARIABILITY AND TIME LAGS}

Models that were proposed to explain temporal properties of accreting black
holes can be generally  divided  into a few groups.  Phenomenological  shot
noise models belong to the first group.  Here, any  dependence  of shots on
energy is just postulated.  Such models can successfully  reproduce  energy
dependence of ACF/CCFs, time lags, PDS, coherence function, etc.  (see e.g.
Miyamoto \&  Kitamoto  1989;  Poutanen  2001; and Fig.  6).  However,  they
leave  unanswered  the question of the origin of spectral  variability.  In
another  class of models,  the  variability  is related to the  Comptonized
component.  Here either seed soft photons are  considered  as the source of
variability with constant  properties of a hot Comptonizing cloud (Kazanas,
Hua, \&  Titarchuk  1997;  B\"ottcher  \& Liang 1998; Hua,  Kazanas, \& Cui
1999; B\"ottcher \& Liang 1999), or the hot cloud itself is responsible for
variability  due to changes of the energy  dissipation  rate  (Poutanen  \&
Fabian 1999; Malzac \& Jourdain 2000) or  propagation  of waves (Kato 1989;
Misra 2000).  The third group  considers  the impact of Compton  reflection
from the accretion disk (e.g.  Vikhlinin 2000;  Poutanen, in  preparation).
Below, I briefly review the proposed  models and  constraints on the models
from observations.

\subsection{Shot Noise Models}

One can attempt to interpret  the observed  ACFs, CCFs and the average shot
profiles in terms of a simple shot noise model  (Terrell  1972).  Since the
CCFs peak at a lag $\la 2$ ms, the shots at different  energies also should
reach maxima within 2 ms from each other.  The energy dependence of the ACF
requires the shots at higher energies to be shorter.  Based on the shape of
the CCF, Maccarone et al.  (2000) (see also Miyamoto \& Kitamoto 1989) also
argued that shot's rise time scale decreases at high energies and the decay
times at  different  energies  are very  close to one  another  or are much
smaller than the rise time scales.  The energy  dependence of the rise time
then produces the hard time lags and the asymmetry of the CCF.

For a broad  distribution  of shot time  scales,  $\tau$,  one comes to the
modified  shot  noise  models  (see,  e.g.,   Lochner  et  al.  1991).  The
properties of the shot behavior at different  energies can be summarized as
follows.  In order to  achieve  high  values  of the  CCFs,  the  shots  at
different  energies  should be perfectly  synchronized.  The observed  hard
lags can be produced  only if hard shots are delayed  relative  to the soft
ones.  This delay is  approximately  a  constant  fraction  of $\tau$  (see
Fig.~6) and should grow as $\ln (E_h/E_s)$  with energy, since the observed
time  lags  are  approximately   inversely   proportional  to  the  Fourier
frequency,  $\delta  t(f)\propto  f^{-1}$,  and grow  logarithmically  with
energy.  And finally, since ACF's width  decreases  with energy, hard shots
should be narrower than soft shots.

\subsection{Comptonization Models}

In the  process  of  Comptonization,  soft  photons  are  scattered  by hot
electrons and gain energy.  Harder  photons are produced in a larger number
of  scattering,  and they  spend on  average  more  time in the hot  cloud.
Therefore, hard photons lag behind soft photons.  The characteristic  delay
between  photons of energies  $E_h$ and $E_s$ for a cloud of size $R$ and 
temperature $\Theta=kT_e/m_ec^2$ is (Sunyaev \& Titarchuk 1980; Payne 1980)  
\be  
t_c= \frac{R}{c(1+\taut)} \frac{\ln (E_h/E_s)}{\ln [1+4\Theta(1+4\Theta)]}.  
\ee  
Comptonization in a uniform electron cloud produces  frequency  independent
time lags  (Miyamoto et al.  1988).  In order to reproduce  large values of
lags,  their  $f^{-1}$  dependence,  and their  logarithmic  dependence  on
energy,  Kazanas et al.  (1997)  proposed an extended  corona  model with a
$r^{-1}$ radial density  distribution and a few light seconds in size.  The
long lags at low Fourier  frequencies  are attributed to photons  traveling
over large  radii,  while the  shorter  time lags at high  frequencies  are
produced  in the  central  small  core of the  cloud.  The  model  predicts
reduced  variability of the higher energy photons at high  frequencies.  On
the other  hand,  the PDS of Cyg X-1  hardens  with  energy  (Nowak  et al.
1999a), i.e.  the high energy photons are more variable.  Using time domain
analysis,  Maccarone et al.  (2000)  showed that the extended  corona model
predicts  the ACF's width to grow with  energy (the 30 keV ACF is ten times
broader  than the 3 keV ACF),  while  the  observations  show  that the ACF
becomes  narrower with energy (see Fig.~3).  They concluded  that any model
where the time lags are the result of light travel delays can be ruled out.
One should note that this, of course, does not rule out Comptonization as a
mechanism for the X-ray production.

\subsection{Magnetic Flares}

The observed spectrum can be generated by magnetic flares on the surface of
the cold accretion  disk (see e.g.  Haardt,  Maraschi, \& Ghisellini  1994;
Stern et al.  1995; Poutanen \& Svensson 1996;  Beloborodov  1999a,b).  The
magnetic fields amplified in the accretion disk and expelled from it due to
the Parker instability  elevate to the corona releasing magnetic energy and
heating the corona.  The flare time scales are probably of the order of the
Keplerian  time scale at a given  distance from the central black hole.  In
the model by Poutanen \& Fabian (1999),  changes in the energy  dissipation
rate and in the  geometry of the flare (e.g.  the  distance  from the disk)
produce soft-to-hard spectral evolution which is the cause of the hard time
lags.  A broad  distribution  of the  flare  time  scale  $\tau$  (as for a
modified  shot  noise  model)  assures  that the time  lags  are  inversely
proportional to the Fourier frequency $\delta t(f)\propto \tau \sim 1/(2\pi
f)$ (see Fig.~6).

This  model  reproduces  well the time  lags  observed  in Cyg X-1.  Energy
dependence  of the  ACF/CCFs  can be  accounted  for  only  if  the  energy
dissipation  rate in flares rises slower than it decays  (Maccarone  et al.
2000).  When the flare  time  scale is  comparable  to the  light  emission
region  crossing  time,  $R/c$, the model  predicts  soft time lags at high
frequencies $f \ga 1/(20 R/c)$ independently of the energy dissipation rate
profile  (Malzac \& Jourdain  2000; see Fig.~7).  This property can be used
to put constraints on the size of the emission region.

\begin{figure}[t]
\begin{minipage}{8.cm}
\includegraphics[width=65mm]{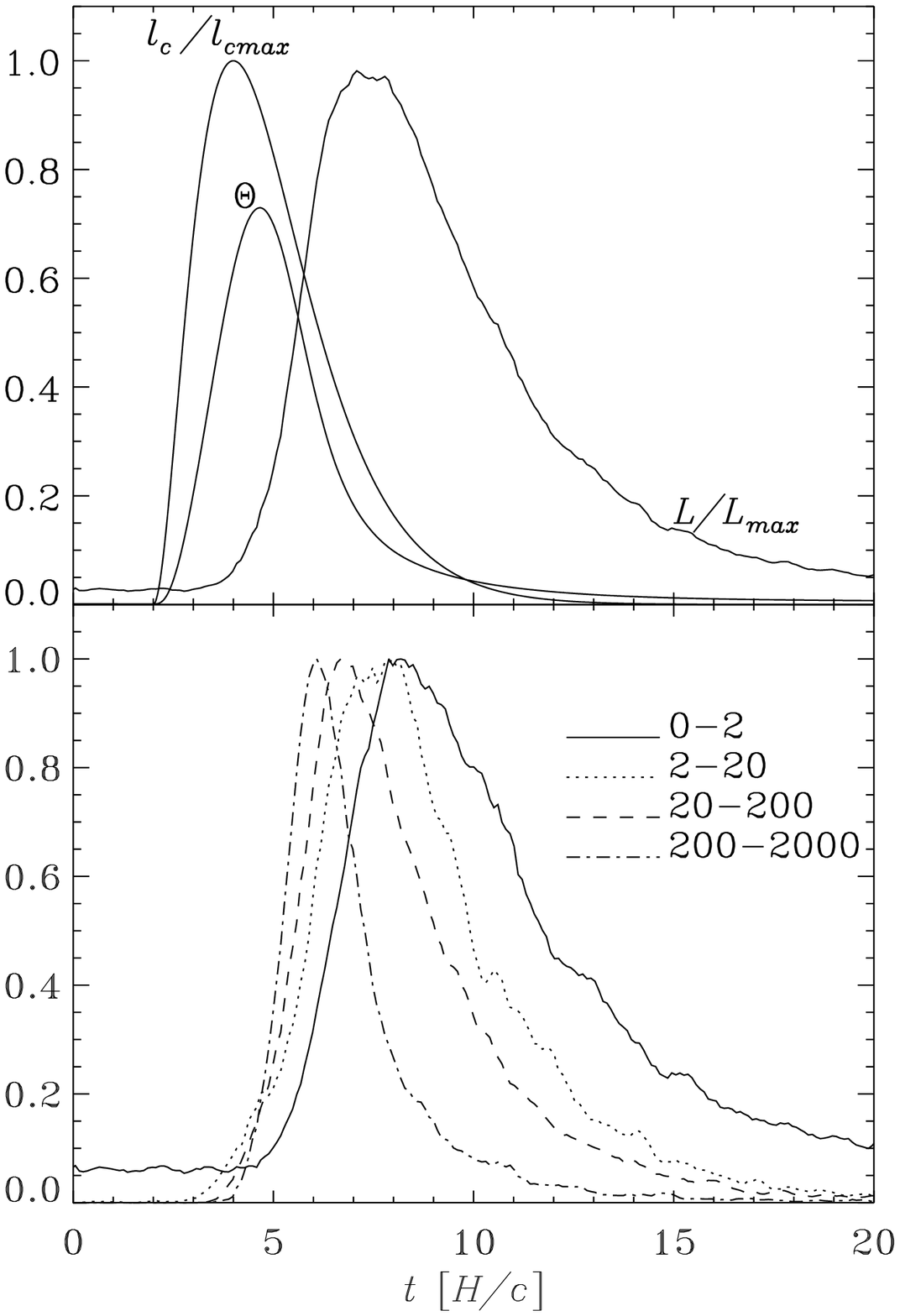}
\end{minipage}
\begin{minipage}{10.cm}
{\sf Fig. 7.    
Spectral  evolution of a strong flare (in a slab-corona geometry).  
Upper panel shows the  evolution of
the dissipation rate $l_c$, electron temperature  $\Theta$, and the emitted
luminosity $L$ (for a constant  $\taut=0.4$).  
Lower panel shows time evolution
of the observed flux in different energy  intervals.  The time scale of the
flare is about the slab  light  crossing  time,  $H/c$.  The seed
photon flux is dominated by reprocessed  photons.  In such  conditions, the
spectrum is hard in the beginning of the flare due to photon starvation (it
takes a few $H/c$ to produce soft photons).  Spectrum softens in the end of
the flare.  Independently of the energy  dissipation temporal evolution,
softer  photons escape on average  after harder ones.  
Adapted  from Malzac \& Jourdain (2000).
}
\end{minipage}
\end{figure}

\subsection{Cold Blobs in Hot Disk}

One of the viable model for the X-ray  emission is the hot inner disk model
(e.g.  Esin et al.  1998).  The soft  photons can be  produced  by external
cold disk or by cold blobs  inside the hot  material.  B\"ottcher  \& Liang
(1999)  considered cold blobs drifting  towards the black hole.  When blobs
enter the innermost hottest region of the disk the radiation was assumed to
increase due to the influx of the soft seed photons.  The overall  spectrum
then  hardens.  The  spectral  evolution  results  in hard time  lags.  The
problem  is that if the  energy  dissipation  rate in the hot disk does not
vary, then an  increase  of the soft  photon  flux  would  lead to a softer
spectrum (not a harder one).  In order to save the model, one has to assume
that the  inward  drift of cool  blobs is  perfectly  correlated  with  the
increase of the energy  dissipation in the hot disk.  It remains to be seen
whether such a requirement is physically realistic and the model can indeed
fit the data.

\newpage

\subsection{Waves in the Disk}

Several papers discussed  models where hard time lags are produced by waves
propagating through the accretion disk from the outer, cooler region to the
inner, hotter region (e.g.,  Miyamoto et al.  1988; Kato 1989; Nowak et al.
1999c).  Frequency  dependent  time lags appear due to  dispersion  of wave
velocities.  Most  of the  models  do not  however  specify  the  radiative
processes  responsible  for the  X-ray  emission.  Recently,  Misra  (2000)
proposed  that   Comptonization   in  a``transition   disk''  with  varying
temperature  could  be  responsible  for  that  emission.  In  that  model,
emission  at 30 keV is  delayed  by $\sim  0.015$ s from the peak at 3 keV.
Maccarone et al.  (2000)  argued  that, in such a case, the CCF between the
30 keV and 3 keV photons would peak at that lag strongly  contradicting the
data for Cyg X-1.  Since the signal at  different  energies is  produced in
physically  separated regions, it is very difficult to understand how it is
possible to keep very high  coherence of the signals at different  energies
(see e.g.  Nowak et al.  1999a).


\subsection{Impact of Compton Reflection on Temporal Characteristics}

Observations show that accretion disks in X-ray binaries are  geometrically
thick at the outer edge (White \& Holt 1982; Mason \& Cordova 1982; Vrtilek
et al.  1990; Hynes et al.  1998; see  Verbunt  1999 for a recent  review).
The height-to-radius  ratio, $H/R$, at the outer edge of the disk can be as
large as 0.15-0.5.  This is much larger than is predicted  by the  standard
accretion  disk  theory.  A large  fraction of the X-rays  from the central
source is intercepted by the disk and some are reflected.  These  (Compton)
reflected  photons  are  significantly   delayed  relative  to  the  direct
emission.  Since  Compton  reflection  spectrum  is very  hard,  at  higher
energies a larger  fraction of photons is delayed.  This produces hard time
lags (see below).

\begin{figure}[t]
\bc
\begin{minipage}{17.cm}
\includegraphics[width=160mm,height=130mm]{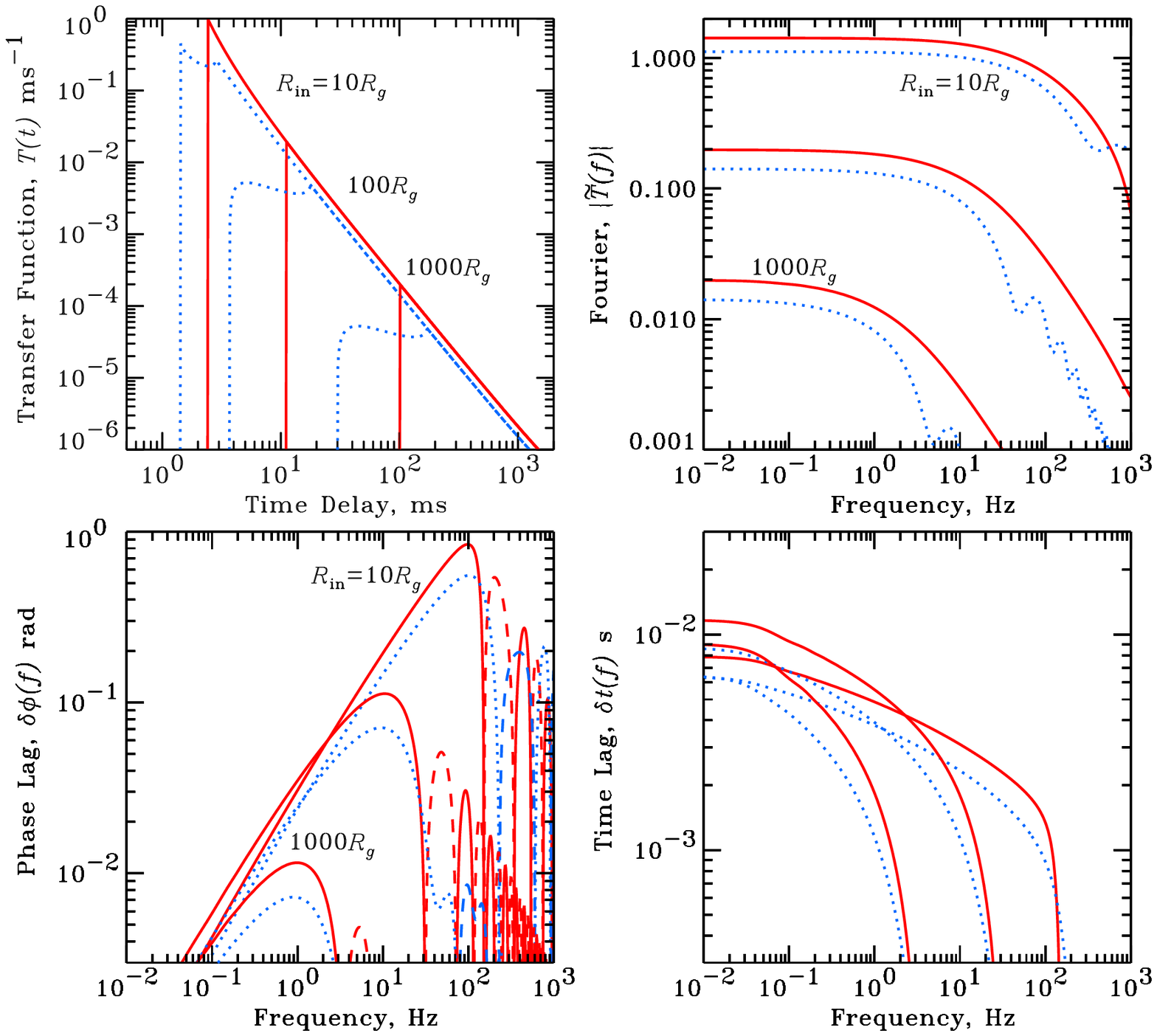}
\end{minipage}
\ec
{\sf Fig. 8. Response and time lags produced by reflection 
off the flat disk with the central hole of different radii $R_{\rm in}$. 
Solid curves are for inclinations  $i=0^{\deg}$ and dotted curves
are for $i=45^{\deg}$. Dashed curves give negative lags.
Isotropic flare is at the center at elevation $h_{\rm fl}=10R_g$. 
Time and frequency units correspond to the $10M_{\odot}$ black hole
($R_g=2GM/c^2=30$ km). Albedos $a_s=0, a_h=1$ are assumed. 
The angular distribution of reflection is $2\cos i$.}
\end{figure}


The  reflection  spectrum  as  a  function  of  time,  $I_r(E,t)$,  can  be
represented as a convolution of the  direct/intrinsic  spectrum,  $I(E,t)$,
with the  transfer  function,  $T(t)$  (normalized  to the  total  observed
amplitude of reflection), describing the time
delay and Green's  function,  $G(E,E')$,  describing the process of Compton
downscattering and photoelectric absorption,
\be\label{eq:ir}
I_r(E,t)= \int_{-\infty}^{t} T(t-t') d t'
\int_{E}^{\infty} G(E,E') I(E',t') d E' .
\ee
This equation can be
simplified if one   neglects the process of Compton  down-scattering (i.e.
for  low  energy  X-ray  photons,  $E\la  10$  keV),  and  approximate  the
redistribution  function by a delta-function,  $G(E,E')=a(E)  \delta(E-E')$
(here $a(E)$ is the reflection albedo).  Then
\be
I_r(E,t)=a(E) \int_{-\infty}^{t} T(t-t') I(E,t') d t' .
\ee
Let us consider  light  curves in the two energy bands   
which are composed of the direct  radiation  from the X-ray source plus the
reflected radiation:
\be
S(t)=S_d(t)+a_s \int_{-\infty}^{t} T(t-t') S(t') dt'  , \quad
H(t)=H_d(t)+a_h \int_{-\infty}^{t} T(t-t') H(t') dt' .
\ee
Here $a_s$ and $a_h$   are the respective X-ray albedos.  
The Fourier transforms are:
\be
\tS(f)=\tS_d(f) [1+a_s \tT(f)], \quad
\tH(f)=\tH_d(f) [1+a_h \tT(f)].
\ee
The cross-spectrum
\be
\tS^*(f) \tH(f) =
\tS^*_d(f) \tH_d(f) \left[ 1+a_s a_h |\tT(f)|^2+ 
a_s \tT^*(f)  +a_h\tT(f) \right]  
= |C_d(f)| e^{i\delta\varphi_d(f)} |C_r(f)| e^{i\delta\varphi_r(f)},
\ee
where $C_d(f)$ and $\delta\varphi_{d}(f)$ are 
the cross-spectrum  and the  phase  lag of the direct radiation.  The
reflection   introduces into the signal an additional phase lag:
\be
\tan \delta\varphi_r(f)= (a_h-a_s)\Im \tT(f) \left/
\left[ 1+(a_s+a_h)\Re \tT(f)  + a_sa_h|\tT(f)|^2 \right] . \right.  
\ee
For illustration, let us consider $a_s\ll 1$ (i.e. $E_s\la 2$  keV), and 
specify a simple exponential transfer function, 
$T(t)=r_1\exp\{-(t-\tau_1)/\tau_1\}/\tau_1$ for  $t>\tau_1$.
After trivial calculations one gets: 
\be
\tan \delta\varphi_r(f)=  a_hr_1 (x \cos x+ \sin x) \left/
\left[ 1+x^2+a_hr_1 (\cos x- x \sin x)\right], \right. \quad x=2\pi f \tau_1.
\ee

\bc
\begin{minipage}{11cm}
\includegraphics[width=100mm]{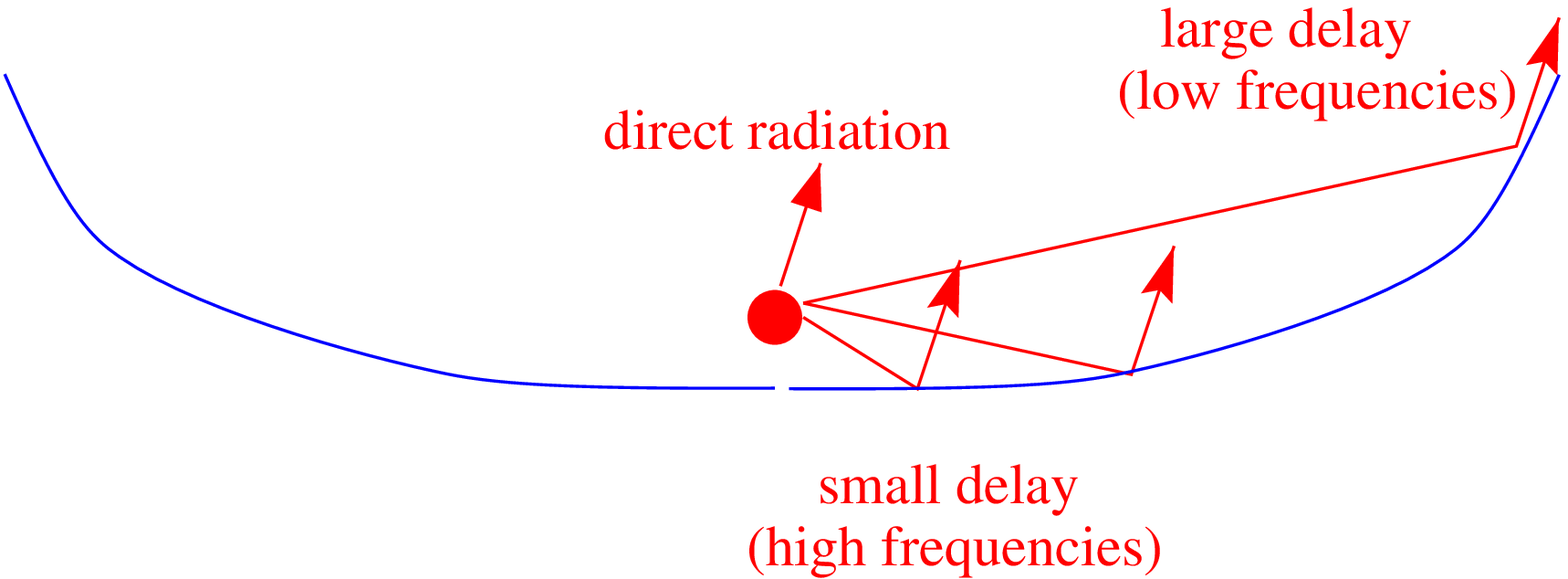}
\end{minipage}
\ec
{\sf Fig. 9. Cartoon of the reflection from flared disk. }

\mbox{ }

\bc   
\begin{minipage}{17.cm}
\includegraphics[width=160mm,height=130mm]{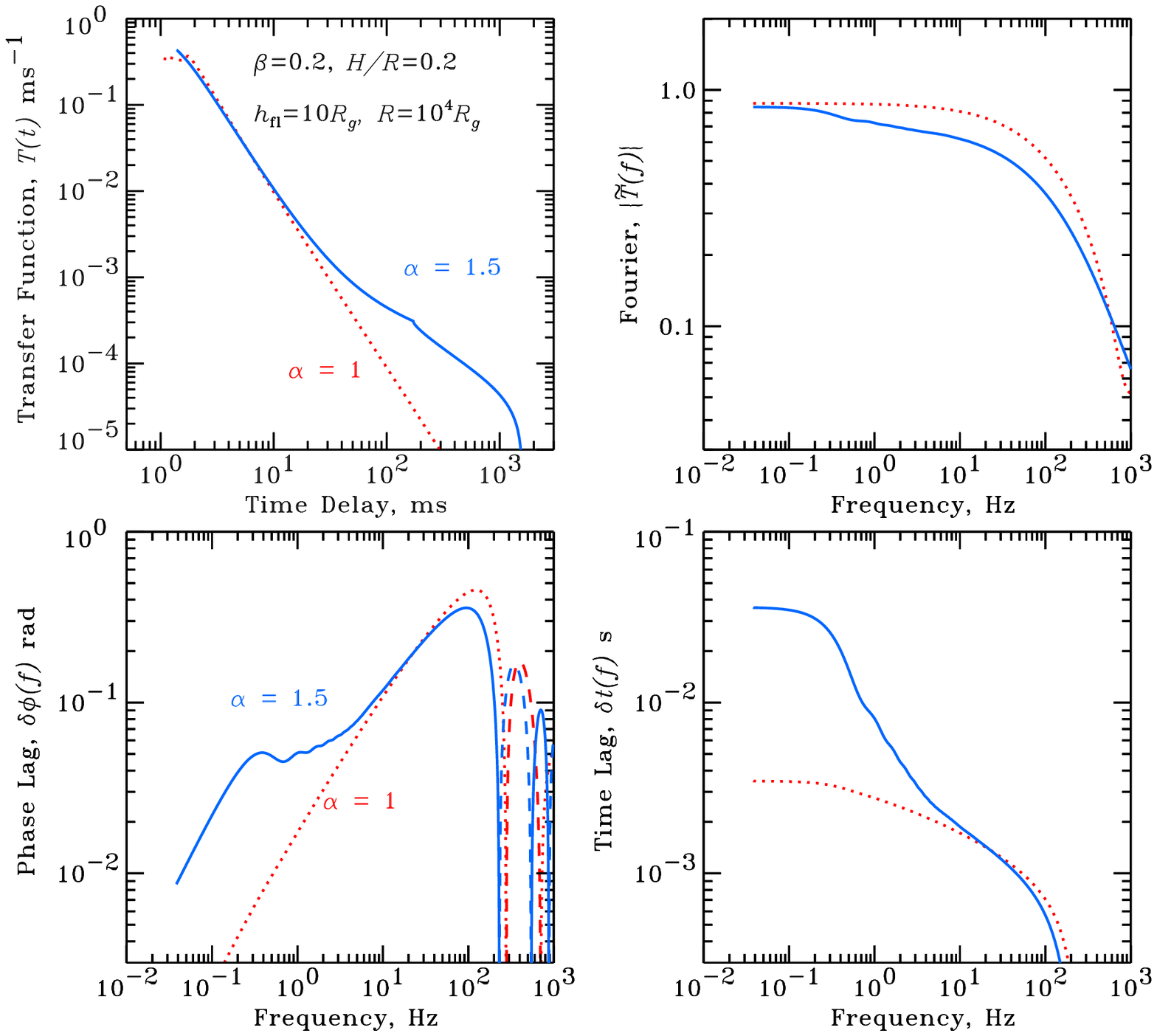} 
\end{minipage} 
\ec
{\sf Fig.  10.  
Response  of the  flared  disk and time  lags.  Flare is at the  center  at
elevation  $h_{\rm fl}=10R_g=300$  km (here it is assumed that 
the inner disk radius is 0); the size of the disk $R=10^4R_g=1$ light sec;
height to radius ratio of the disk $H/R=0.2$;  bulk
velocity of the plasma $\beta=v/c=0.2$ and inclination  $i=45^{\deg}$.  
Height of the disk  scales  with  radius  as  $h(r)\propto  r^\alpha$.  
Solid  curves correspond to $\alpha=1.5$ and dotted curves to $\alpha=1$.
Dashed curves give negative lags.
}

\mbox{ }

\noindent
The phase lag grows as $\delta\varphi_r(f)\approx  2a_hr_1x/(1+a_hr_1)$ for
$x\ll  1$,  and  reaches  the  maximum  $\tan\varphi_{r,\max,1}\approx  0.7
a_hr_1/(1-0.15a_hr_1)$  at  $x\approx  1$.  One should  note that for small
$a_hr_1$,  the energy  dependence  of the phase lag is exactly  the same as
that for the albedo $a_h$.  
If the transfer function is a sum of two exponentials
described by $\tau_1, r_1$ and $\tau_2, r_2$,  respectively (let $\tau_1\ll
\tau_2$),  the phase  lag has two  prominent  maxima  at  $f_1\sim  1/(2\pi
\tau_1)$ and $f_2\sim  1/(2\pi  \tau_2)$.  The value of the high  frequency
maximum at $f_1$  stays the same,  while the new low  frequency  maximum is
$\tan\varphi_{r,\max,2}\approx 0.7 a_hr_2/[1+a_h(r_1-0.15r_2)]$.


Let us consider a specific  physical  situation  where the X-rays  produced
close to the black hole (e.g.  the hot disk) are reflected from a flat cold
disk of a varying inner radius,  $R_{\rm in}$  (Poutanen et al.  1997; Esin
et al.  1998).  For simplicity, we consider a point source in the center at
elevation,  $h_{\rm fl}$.  The transfer  functions and time/phase  lags are
shown  in  Figure~8  (see  also  Gilfanov  et  al.  2000).  An   analytical
expression  for the  transfer  function  can be  easily  obtained  for  the
inclination $i=0$, $T(t)=2ch_{\rm  fl}/(ct-h_{\rm fl})^2$, where $ct>h_{\rm
fl}+\sqrt{h_{\rm  fl}^2+R_{\rm  in}^2}$  (a  $2\cos  i$  dependence  of the
amplitude reflection on inclination is assumed).  Reflection works as a low
pass filter,  therefore the time lags appear only at low  frequencies.  For
small inner radius,  $R_{\rm  in}=10R_g$,  the  amplitude of  reflection is
significant and the maximum phase lag  corresponds to the light travel time
from the  flare to the  cold  disk  edge.  For  larger  $R_{\rm  in}$,  the
reflection is small and the time lags do not grow in spite of the fact that
the characteristic distance to the reflector increases.

An  interesting  change in the time lag  behavior  occurs  when the disk is
flared (see Fig.~9).  Then the outer parts of the disk occupy a significant
solid angle as viewed from an X-ray source in the center.  The  intercepted
photons  can be delayed by a few  seconds  (in GBHs).  The X-rays  emission
regions  (magnetic  flares) can be elevated above the disk and  distributed
all over its central part.  For  simplicity, we consider a ``filled''  disk
(i.e.  zero inner radius) and a point source elevated by $h_{\rm fl}=10R_g$
above the  center of the  disk.  Any  small  offset  from the disk  axis is
negligible   (see  e.g.  Gilfanov  et  al.  2000).  The  response  and  the
amplitude  of time lags depend on the  angular  distribution  of the direct
radiation  from the X-ray flare.  If plasmas in flares has bulk motion away
from the disk the  radiation is beamed  towards the observer  reducing  the
apparent  amplitude of reflection  (and time lags).  Non-zero bulk velocity
is required to explain the hard X-ray spectra and small reflection observed
in GBHs (Beloborodov  1999a,b; Malzac, Beloborodov, \& Poutanen 2001).  The
response of the disk can be represented approximately as a sum of responses
from an  infinite  plane and from the outer edge of the disk.  The  results
then can easily be  understood  from a trivial  example of two  exponential
transfer  functions (see above).  Numerically  computed transfer  functions
and time lags are presented in Figure~10.  For a  $h(r)\propto  r^{\alpha}$
profile of the disk, $\alpha>1$  produces a secondary  maximum in the phase
lag  spectrum  corresponding  to the outer  edge of the  disk.  It could be
responsible for the break in the time-lag Fourier spectrum observed in GBHs
(see Fig.~6a).

Reflection is most probably not the main mechanism  producing lags in GBHs,
otherwise  their  energy  dependence  would  be the  same  as  that  of the
reflection  albedo  having  local  maximum  around Fe  K$\alpha$  line (not
observed).  The  reflection,  however,  {\it is} observed in the spectra of
GBHs and  therefore  it must have an  impact  onto the  timing  properties.
Determining the fraction of time lags produced by reflection is a matter of
future work.

\section*{SUMMARY}

There are several  competing  models for the X-ray  production in accreting
black holes (see  Beloborodov,  this  volume).  Most of these models make a
number  of ad  hoc  assumptions  that  are  difficult  to  check.  Temporal
variability  provides  independent tests for the models.  Already now there
are a number of important  consequencies from comparing model prediction to
the data.  For example, the  mysterious  hard time lags cannot  result only
from light  travel  effects in the  Comptonizing  cloud.  In that case, the
ACF's width would  increase  with photon  energy  which is opposite  to the
observed  behavior.  Strong  constraints  on the  temporal  behavior of the
energy  dissipation  rate in flares can be obtained  from the average  shot
profiles,  CCFs,  and the  Fourier-frequency  dependent  lags.  Most of the
models  concentrate on the properties of the Comptonizing  radiation, while
the  Compton  reflection  can (and  should)  have a  strong  impact  on the
observed variability properties as shown here.

Unfortunately,  basically  all the  models  that are  designed  to  explain
temporal  characteristics  are  still  phenomenological  and they are often
based (as spectral models) on a number of questionable assumptions.  We are
making the first steps in  designing  a model that would be able (at least)
not to contradict the  observational  facts.  Concluding, I just would like
to point out that  real  objects  (such as Cyg X-1, for  example)  are very
complex.  Thus  one can  hardly  expect  that a model  that  satisfies  the
observational facts should be very simple.

\section*{ACKNOWLEDGEMENTS}

This research was   supported by the Swedish  Natural  Science  Research
Council and the Anna-Greta and Holger Crafoord Fund.

\end{document}